\documentstyle[12pt,epsfig]{article}
\newcommand{\lef}{(1-\gamma_5)}
\newcommand{\rig}{(1+\gamma_5)}

\newcommand{\cms}{\frac{\alpha_s^2}{216 M^2_{sq}}}
\newcommand{\ded}{(\delta^d_{12})}
\newcommand{\MK}{M_K f_K^2}
\newcommand{\Mfrac}{\left(\frac{M_K}{m_s+m_d}\right)^2}
\newcommand{\Mfracmu}{\left(\frac{M_K}{m_s(\mu)+m_d(\mu)}\right)^2}

\newcommand{\msg}{m_{\tilde{g}}}
\renewcommand{\a}{\alpha}
\renewcommand{\b}{\beta}

\renewcommand{\d}{\delta}

\newcommand{\g}{\gamma}
\newcommand{\eps}{\epsilon}
\newcommand{\bea}{\begin{eqnarray}}
\newcommand{\eea}{\end{eqnarray}}
\newcommand{\beq}{\begin{equation}}
\newcommand{\eeq}{\end{equation}}
\newcommand{\nn}{\nonumber}
\newcommand{\fr}{\frac}
\newcommand{\hl}{\hline}
\begin{document}

\begin{titlepage}
\begin{flushright}
ROME1-1216/98\\
\end{flushright}
\vskip 1cm
\centerline{\LARGE 
The Supersymmetric Flavor Problem for}
\centerline{\LARGE Heavy First-Two Generation Scalars at}
\centerline{\LARGE  Next-to-Leading Order }
\vskip 1cm
\centerline{R. Contino and I. Scimemi}
\vskip 0.5cm
\centerline{ Dip. di Fisica,
Universit\`a degli Studi di Roma ``La Sapienza" and}
\centerline{INFN, Sezione di Roma, P.le A. Moro 2, 00185 Roma, Italy. }
\begin{abstract}
We  analyze in detail the constraints on SUSY-model
parameters obtained from $\rm K-\overline{K}$ mixing in the hypothesis
of a splitted SUSY spectrum. FCNC contributions  from gluino-squark-quark
interactions  are studied in the so-called mass insertion approximation.
We  present boundaries on mass insertions and on SUSY mass scales.
We improve previous results by including the 
NLO-QCD corrections to $\Delta S=2$ effective Hamiltonian and the complete set of 
B-parameters for the evaluation of hadronic matrix elements. A full set of magic-numbers,
that can be used for further analyses of these models, is also given.
We find that the inclusion of NLO-QCD corrections and the B-parameters
change the results obtained at LO and in the Vacuum Insertion Approximation
by an amount of about $25-35\%$.
\end{abstract}
\date{}
\end{titlepage}
\newpage

\section{Introduction}

It is well known that SuperSymmetry (SUSY) introduces many new sources of
Flavor Changing Neutral Currents (FCNC) which give strong constraints on the
construction of extensions of the Standard Model (SM).

A common feature of these models is that FCNC effects are induced by SUSY breaking
parameters that mix different flavors.
In the literature several ideas have been proposed in order to suppress unwanted FCNC
effects. For instance, in models where SUSY breaking is induced by gauge
interactions~\cite{mdin} SUSY breaking parameters are flavor blind  or they are dominated
by the dilaton multiplet of string theory~\cite{kapl}.
Alternatively, flavor symmetries are used to provide either a sufficient degeneracy
between the first-two generation of sfermions~\cite{leig} or alignment between quark
and squark mass matrices~\cite{ynir}.

Here we want to investigate  the hypothesis that the  average squark mass
of the first-two generations is much higher then the rest of the spectrum of
(s-)particles~\cite{dine,dimo,poma,cohe}. Throughout the paper we indicate  the average
mass of the heavy scalar squarks as $M_{sq}$ and  the
typical mass scale of gauginos and 
of the other light sparticles as  $\msg$. Small Yukawa couplings of the first-two
generations of scalars to the Higgs doublets, together with masses of the rest of the
supersymmetric spectrum close to the weak scale allow  a natural electroweak symmetry
breaking (EWSB). 
This scenario has very interesting phenomenological
signatures~\cite{cohe} and can  be easily realized in  string theory~\cite{bine}.

We consider gluino-squark-mediated FCNC  contributions
to $\Delta M_K$ and $\eps_K$ in the neutral $K$-system.
The  effect of the most general squark mass matrix for this class of models is studied.
In some cases further   restrictions on the squark masses are required
and other contributions can be more important. 
In particular chargino-squark-quark interactions should be also considered.
We postpone a discussion with the inclusion of these effects to a subsequent work.

We work in the so-called mass insertion  approximation~\cite{hall}.
In this  framework one chooses a basis for  fermions and sfermions states
where all the couplings of these particles to neutral gauginos are flavor diagonal and FC
effects are shown by the non-diagonality of  sfermion propagators.
The pattern of flavor change, for the K-system, is given by the ratio 
\beq
(\delta^{d}_{ij})_{AB}= 
\fr{(m^{\tilde d }_{ij})^2_{AB}}{M_{sq}^2} \ ,
\eeq
where $ (m^{\tilde d }_{ij})^2_{AB}$  are the off-diagonal elements of the
$\tilde d $ mass squared matrix that mixes flavor $i$, $j$ for both left- and
right-handed scalars ($A,B=$Left, Right), see {\it e.g.}~\cite{gabb}.
The sfermion propagators are expanded as a series in terms of the $\delta$'s and the
contribution of the first term of this expansion is considered.

The supersymmetric flavor problem consists in  building viable models in which FCNC
are  suppressed without requiring  excessive fine tuning of the parameters. 

In models with a splitted spectrum of s-particles, in which the average mass of the
lightest ($m_{\tilde g}$) is in the electroweak or TeV region, two scenarios are
possible:
\begin{enumerate}
\item
for reasonable values of $M_{sq}$, the suppression of FCNC requires small $\delta$'s
values. Thus, by fixing $M_{sq}$, one can find constraints on $\delta$'s, see
refs.~\cite{gabb,bagg} and for a very recent NLO analysis ref.~\cite{ciuc2};
\item
for natural values of the $\delta$'s, say ${\cal{O}}(1)$ or order
of the Cabibbo angle, ${\cal{O}}(0.22)$, one finds that the only way to get rid of
unwanted FCNC effects, is by having  the squarks of the first-two generations heavy
enough. Thus, by fixing the $\d$'s, one can find constraints on the
minimal values for $M_{sq}$. 
Large values of $M_{sq}$, however, induce large values for
the GUT masses of the third generation of squarks via Renormalization Group Equations
(RGE). Consequently there can be fine-tuning  problems for the $Z$-boson mass.
We study  this issue in sec.~\ref{sec:spec}.
This point of view was adopted in  refs.~\cite{arka,agas}. 
\end{enumerate}
In the past, several phenomenological analyses were carried out,
which relied on some approximations. For instance, the work of ref.~\cite{arka} does
not include QCD radiative corrections and makes use of Vacuum Insertion Approximation
(VIA) for the evaluation of hadronic matrix elements.
Leading order QCD corrections to the evolution of  Wilson coefficients were,
instead, considered in the papers of refs.\cite{bagg,agas}.
These authors found that QCD corrections are extremely important.
For example in \cite{agas} the lower bound on the heavy squark mass
is increased by roughly a factor three. 

In this work we discuss both the cited scenarios and improve previous analyses including 
the Next-to-Leading Order (NLO) QCD corrections to the most general $\Delta F=2$
effective Hamiltonian~\cite{ciuc} and the lattice calculation of all the
B-parameters appearing in the $K-\bar K$ mixing matrix elements
that have been recently computed~\cite{allt}. We find this very interesting for several
reasons. First of all we find that the  inclusion of these effects leads to sizeable
deviation from the previous computations.
The results obtained using only LO-QCD corrections and Vacuum Insertion
Approximation are  corrected by about $25-35\%$.
Furthermore,  the uncertainties  of the final result due to its dependence on the scale
at which hadronic matrix elements and quark masses are evaluated is much reduced.

Predictions for any model can be tested using the so-called magic numbers
we provide. These numbers allow to obtain the coefficient functions
at any low energy scale
once the matching conditions are given at a higher energy scale.
The magic-numbers  will be useful, {\it e.g.},
when a complete NLO analysis of SUSY contributions to $\Delta F=2$
processes (which should include also chargino exchange effects)
will be  implemented in the future.

A complete NLO calculation should be comprehensive also of the 
${\cal{O}}(\a_s)$ corrections
to the Wilson coefficients at the scale of the SUSY masses running in the loops.
So far, we miss this piece of information for gluino-squark
contributions\footnote{The matching conditions for charged-Higgs and chargino 
contributions  have been recently computed~\cite{krau}.}.
We can argue the smallness of these corrections  from the smallness of $\a_s$ at such
scales.
This uncertainty can be removed only by a direct computation.

The paper is organized as follows. In sec.~\ref{sec:heff} we introduce the formalism
concerning the operator basis, the Wilson coefficients and the Renormalization Group
Equations (RGE). In sec.~\ref{sec:deltas} constraints on the $\d$'s are derived.
The problem of consistency of the squark spectrum  for
given entries of the $\d$'s  is considered in sec.~\ref{sec:spec}.
 Finally  our conclusions are written
in sec.~\ref{sec:conc} and  all the magic-numbers are given in the appendix.

\section{Effective Hamiltonian and hadronic matrix elements}

\label{sec:heff}
In this section we describe the framework in which the basic calculations
have been performed. We follow the discussion of ref.~\cite{bagg} in the case
$M_{sq}\gg m_{\tilde g}$. Throughout the paper (unless otherwise explicitly specified),
we assume that the average mass of gluinos and of the squarks of the third generation
are of the same order of magnitude.

The three steps needed to use the Operator Product Expansion (OPE) 
(matching of the effective theory, perturbative evolution of the coefficients and
evaluation of hadronic matrix elements) are treated in detail in the following 
subsections.

\subsection{Operator basis and matching  of the effective theory  }

In order to  apply  the OPE  one has to calculate coefficients and operators of the
effective theory. One first integrates out the heavy scalars of the first-two generations
at the scale $M_{sq}$. This  step produces $\Delta S=1$ (of the form 
$\bar{d} \tilde{g}\bar{\tilde g} s$) as well as $\Delta S=2 $ operators, 
at the same order $1/M_{sq}^2$. When also gluinos  are integrated out  at $\msg$,
$\Delta S=1$ operators generate  $\Delta S=2$ contributions that are proportional to
$m_{\tilde g}^2/M_{sq}^4$, and so can be neglected.

The final basis of operators is:
\bea 
Q_1 &=& \bar d^\alpha \gamma_\mu \lef s^\alpha \ \bar d^\beta \gamma_\mu \lef 
s^\beta\, , \nn \\ 
Q_2 &=& \bar d^\alpha \lef s^\alpha \ \bar d^\beta \lef s^\beta \, , \nn \\ \nn
Q_3&=& \bar d^\alpha \lef s^\beta \ \bar d^\beta \lef  s^\alpha \, , \\ 
Q_4 &=& \bar d^\alpha \lef s^\alpha \ \bar d^\beta \rig s^\beta \, , \nn \\ 
Q_5&=& \bar d^\alpha \lef s^\beta \ \bar d^\beta \rig s^\alpha\, ,  \label{eq:susy} 
\eea 
together with  operators $\tilde Q_{1,2,3}$ which can be obtained from $Q_{1,2,3}$ by
the exchange $\lef \leftrightarrow \rig$.

The Wilson coefficients at the matching scale $M_{sq}$ are
(see {\it e.g.}~\cite{gabb,bagg}):
\bea
\nn
C_1&=&-\cms \left(24 x f_6(x)+66 \tilde{f}_6(x)\right) \ded_{LL}^2, \\ \nn
C_2&=&-\cms 204  f_6(x)\ded_{RL}^2,\\ \nn
C_3&=&\cms 36 x f_6(x)\ded_{RL}^2,\\  \nn
C_4&=&-\cms \left[ \left(504 x f_6(x)-72\tilde{f}_6(x)\right)\ded_{LL}\ded_{RR}\right.
\\ \nn
& &\left. -132 \tilde{f}_6(x)\ded_{LR}\ded_{RL} \right],\\ \nn
C_5&=&-\cms \left[ \left(24 x f_6(x)+120\tilde{f}_6(x)\right)\ded_{LL}\ded_{RR}\right.
\\  & &\left. -180 \tilde{f}_6(x)\ded_{LR}\ded_{RL} \right],
\eea
where $x=(m_{\tilde g}/M_{sq})^2$ and
\bea
f_6(x) &=& \fr{6(1+3 x)\ln{x}+x^3-9 x^2-9 x+17}{6 (x-1)^5}, \nn \\
\tilde{f}_6(x) &=&\fr{6x(1+x)\ln{x}-x^3-9 x^2+9 x+1}{3(x-1)^5}.
\eea
The coefficients for the operators $\tilde Q_{1,2,3}$ are the same as those 
of $Q_{1,2,3}$ with the replacement $L \leftrightarrow R$. The authors of
refs.~\cite{bagg,arka,agas} use the matching coefficients directly in the limit
$x\rightarrow 0$. However, we have contemplated  also the extreme case of
$m_{\tilde g} \sim M_{sq}/2$, so that we  keep the whole expression.
Of course, the value of the coefficients is the same as that of 
refs.~\cite{bagg,arka,agas} in cases where $x\ll 1$.

As we said, NLO-corrections to these coefficients have not  been computed yet.
We assume   they are negligible, in view of  the smallness of $\a_s(M_{sq})$ and of the fact that
similar corrections turned out to be rather small in the SM, the
 two Higgs doublet model, and for the  chargino contribution in the 
constrained MSSM~\cite{krau}. Our effective Hamiltonian is so  affected by a residual
renormalization scheme dependence because of the missing
piece of ${\cal{O}}(\a_s(M_{sq}))$ in the matching.

\subsection{Evolution of Wilson coefficients and running of $\a_s$  }

In order to evolve the Wilson coefficients between $M_{sq}$ and the scale
at which hadronic matrix elements are evaluated ($\mu$=2 GeV), one
has to account for the presence of all particles whose mass is intermediate
between the two scales,  both in the $\b$-function  of $\a_s$ and in the 
Anomalous Dimension Matrix (ADM) of the operators.

For what concerns the former one has~\cite{drtj}:
\bea
\b(\a_s) &=&-\b_0 \a_s^2 - \b_1 \a_s^3 + {\cal{O}}(\a_s^4), \\
\b_0&=&\fr{1}{3} \left(11 N_c -2 n_f -2 N_c n_{\tilde g} -\fr{1}{2} n_{\tilde q}
\right), \\   
\b_1&=& \fr{1}{3} \left(34 N_c^2- \fr{13 N_c^2 -3}{ N_c} n_f- 16 N_c^2 n_{\tilde g} -
\fr{4 N_c^2-3 }{2 N_c} n_{\tilde q}
+ 3 \ \fr{3 N_c^2-1}{2 N_c} n_{\tilde q} n_{\tilde g}  \right),      
\eea
where $N_c=3$ for color SU(3) and $n_f$ is the number of  fermion flavors. 
The terms proportional to  $ n_{\tilde g} $ and $n_{\tilde q }$ represent,
respectively, the gluino and light scalar contributions. $n_{\tilde g}$=1 and
$n_{\tilde q }$=4 when one evolves between $M_{sq}$ and $\msg$ and to
$n_{\tilde g}=n_{\tilde q }$=0 evolving from $\msg$ to a lower mass scale.

In ref.~\cite{ciuc} the Anomalous Dimension Matrix of the operators was computed at NLO.
In that reference, since all SUSY particles are taken to be heavy,
   only  loops with fermions and gluons were considered. 
This result must be modified taking into account that,  from
$M_{sq}$ to $m_{\tilde g}$, also the squarks of the third generation
and gluinos can run in the loops.
As a matter of  fact, for the K-system, light third generation squarks and gluinos can enter
two-loops ADM only via the renormalization of the gluon propagator.
An  explicit calculation shows that the required modification consists in considering
 the ADM of ref.~\cite{ciuc} as a function of
$n_f+N_c \ n_{\tilde g} +n_{\tilde q}/4$ when one evolves between the heavy squark and
gluino mass scales and as a function of  $n_f$  below the latter  scale.
This substitution is no more true if the  squarks of the first-two generations are light too.

The value of the Wilson coefficients at the hadronic scale $\mu=2$ GeV where 
matrix elements are computed can then be easily calculated.
Following ref.~\cite{ciuc} one evolves between two scales according to:
\bea
\vec{C} (\mu)&=&\hat{N}[\mu] \   \hat{U}[\mu,M] \  \hat{N}^{-1}[M] \;
\vec{C} (M),\nn \\ 
\hat{N}[\mu]&=& \hat{1} + \fr{\a_s(\mu)}{4\pi} \ \hat{J}(\mu), \nn \\           
\hat{U}[\mu,M]&=&\left[ \fr{\a_s(M)}{\a_s(\mu)}\right]^{\hat{\g}^{(0)T}/(2 \b_0)},
\eea                 
where $\hat{\g}^{(0)}$ is the LO-ADM and $\vec{C}(\mu)$ are the Wilson coefficients
arranged in a column vector. This formula is correct up to the NLO. $\hat{U}[\mu,M]$
gives the LO evolution already computed in ref.~\cite{bagg} while $\hat{J}$ gives
the NLO corrections calculated in ref.~\cite{ciuc}.
$\hat{J}$ depends both on the number of active particles at the scale $\mu$, and
on the renormalization scheme used for its computation.
We have  used $\hat{J}$ in the same scheme used for the lattice calculation
of hadronic matrix elements, that is the so-called Landau-RI scheme.
In this way the  renormalization scheme dependence of the final result, at the scale at
which hadronic matrix elements are evaluated, cancels out at this perturbative order.
As already been stressed, for a complete scheme independence of our result one should
include also the NLO corrections of the Wilson coefficients at the high matching scale.

We provide here the full set of Wilson coefficients at $\mu$=2 GeV as functions of
$M_{sq}$ and $\msg$ (the so-called magic-numbers).  
We find
\bea
C_i(\mu)&=&\sum_{r,j=1}^{5} \ \left[b_{ij}^{(r)}+ \fr{\a_s(\msg)}{4 \pi}c_{ij}^{(r)}
\right] \ \a_s^{a_r}(\msg) \ C_{j} (\msg), \nn \\
C_i(\msg)&=&\sum_{r,j=1}^{5} \ \left[
d_{ij}^{(r)}+\fr{\a_s(\msg)}{4 \pi}e_{ij}^{(r)} +\fr{\a_s(M_{sq})}{4 \pi}f_{ij}^{(r)}
\right] \ 
\left(\fr{\a_s(M_{sq})}{a_s(\msg)}\right)^{a^\prime_r} C_{j} (M_{sq}).
\label{eq:magic}
\eea
The complete expression of $a_r$, $a^\prime_r$, $b^{(r)}_{ij}$, $\dots$, is given in
the appendix. Eq.~(\ref{eq:magic}) is useful for testing predictions for any model, once
the two scales are fixed. 
The magic numbers for the evolution of $\tilde{C}_{1-3}$ are the same
as the ones for the evolution of $C_{1-3}$.
Eq.~(\ref{eq:magic}) and the formulae of the appendix, can be
 used with B-parameters  evaluated at $\mu=2$ GeV (see eq.~(\ref{eq:bpar})),
in order to determine the contribution
to $\Delta M_K$ and $\epsilon_K$  at NLO in QCD for any model
of new physics in which the new contributions with respect to the SM
originate from the extra heavy particles.
It is sufficient to compute the values of the coefficients
at the matching scales $M_{sq}$ and $m_{\tilde g}$ and put them in
eq.~(\ref{eq:magic}).

\subsection{Hadronic Matrix Elements}

\label{sub:hme}
The hadronic matrix elements of the operators of eq.~(\ref{eq:susy}) in the Vacuum
Insertion Approximation (VIA) are:
\bea
\langle K^0|Q_1|\bar{K}^0\rangle_{VIA}&=&\fr{1}{3}\MK ,\nn \\
\langle K^0|Q_2|\bar{K}^0\rangle_{VIA}&=&-\fr{5}{24}\Mfrac \MK ,\nn \\
\langle K^0|Q_3|\bar{K}^0\rangle_{VIA}&=&\fr{1}{24} \Mfrac \MK ,\nn \\
\langle K^0|Q_4|\bar{K}^0\rangle_{VIA}&=&\left[\fr{1}{24}+\fr{1}{4} \Mfrac\right] 
\MK ,\nn \\
\langle K^0|Q_5|\bar{K}^0\rangle_{VIA}&=&\left[\fr{1}{8}+\fr{1}{12} \Mfrac\right] \MK ,
\eea
where $M_K$ is the mass of the $K$ meson and $m_s$, $m_d$ are the masses of the
$s$ and $d$ quarks respectively. An analogous definition holds for ${\tilde{Q}}_{1,2,3}$.

Hadronic matrix elements  can be evaluated non-perturbatively introducing
B-pa\-ra\-me\-ters, defined as follows:
\bea
\langle K^0|Q_1(\mu)|\bar{K}^0\rangle&=&\fr{1}{3}\MK B_{1}(\mu),\nn \\
\langle K^0|Q_2(\mu)|\bar{K}^0\rangle&=&-\fr{5}{24}\Mfracmu \MK B_{2}(\mu),\nn \\
\langle K^0|Q_3(\mu)|\bar{K}^0\rangle&=&\fr{1}{24} \Mfracmu \MK B_{3}(\mu),\nn \\
\langle K^0|Q_4(\mu)|\bar{K}^0\rangle&=&\fr{1}{4} \Mfracmu \MK B_{4}(\mu),\nn \\
\langle K^0|Q_5(\mu)|\bar{K}^0\rangle&=&\fr{1}{12} \Mfracmu \MK B_{5}(\mu),
\label{eq:bibi}
\eea
where $Q_i(\mu)$ are the operators renormalized at the scale $\mu$. The B-parameters for
${\tilde{Q}}_{1,2,3}(\mu)$ are the same as those of $Q_{1,2,3}(\mu)$.

In the computation of $ B_i$  for the operators 2-5, smaller contributions of higher
order in chiral expansion, coming from axial current, have been neglected.  
A detailed explanation of the reasons of this approximation can be found in
ref.~\cite{allt}. The definition of B-parameters in eq.~(\ref{eq:bibi}) takes explicitly
into account this approximation and  using it the low scale ($\mu$) dependence of the
final result is explicitly canceled in the product of coefficient functions and hadronic
matrix elements.
 
The B-parameter of the first operator is usually addressed as $B_K$ and 
has been extensively studied on the lattice and used in many phenomenological 
applications (see, {\it e.g.}~\cite{shar,lusi}). We have considered its world
average~\cite{shar}. The other $B_i$ have been taken from ref.~\cite{allt}
(for another determination of these $B_i$, calculated with perturbative
renormalization see ref.~\cite{gupta}).

All the B-parameters are evaluated at a scale of 2 GeV in the LRI renormalization scheme:
\bea
B_{1}(\mu=2\ {\rm GeV}) &=&0.60 \pm 0.06,\nn \\
B_{2}(\mu=2\ {\rm GeV}) &=&0.66 \pm 0.04,\nn \\
B_{3}(\mu=2\ {\rm GeV}) &=&1.05 \pm 0.12,\nn \\
B_{4}(\mu=2\ {\rm GeV}) &=&1.03\pm 0.06,\nn \\
B_{5}(\mu=2\ {\rm GeV}) &=&0.73 \pm 0.10.
\label{eq:bpar}
\eea

So far in the literature all phenomenological analyses on this subject have 
used the VIA and  have computed Wilson coefficients and quark masses at a scale variable
between 0.5-1 GeV. We will see this represents in some cases quite a rough approximation.
    
Finally we give in table~(\ref{tab:cost}) all the numerical values of the physical 
constants we have considered. All coupling constants and $\sin^2{\theta_W}(M_Z)$
are meant in the $\overline{MS}$-scheme~\cite{pdg}.
\begin{table}
\begin{center}
\begin{tabular}{||c|c||}\hl
Constants & Values \\
\hl \hl
$\a_{em} (M_Z)$ & 1/127.88 \\ \hl
$\a_s (M_Z)$ & 0.119 \\ \hl
$M_K$ & 497.67 MeV \\ \hl
$f_K$ & 159.8 MeV \\ \hl
$m_d$(2 GeV)&7 MeV\\ \hl
$m_s$(2 GeV)&125 MeV\\ \hl
$m_c$&1.3 GeV\\ \hl
$m_b$&4.3 GeV\\ \hl
$m_t$&175 GeV\\ \hl
$\sin^2{\theta_W}(M_Z) $ & 0.23124 \\
\hl \hl
\end{tabular}
\caption{ Constants used for phenomenological analysis.}
\label{tab:cost}
\end{center}
\end{table}

\section{Constraints on the $\d$'s}

\label{sec:deltas}
We are ready to provide a set of constraints on SUSY variables coming from the $K_L-K_S$
mass difference, $\Delta M_K$ and the CP violating parameter $\eps_K$ defined as
\bea
\Delta M_K &=&2 {\rm{Re}}\langle K^0|H_{\rm{eff}}|\bar{K}^0\rangle, \nn \\
\eps_K &=& \fr{1}{\sqrt{2} \Delta M_K}{\rm{Im}}\langle K^0|H_{\rm{eff}}|\bar{K}^0\rangle.
\eea
The parameter space  is  composed of two real and four complex entries, that
is $M_{sq}$, $m_{\tilde g}$ and $\ded_{LL}$, $\ded_{LR}$, $\ded_{RL}$, $\ded_{RR}$.

Neglecting interference
among different SUSY contributions, we give upper bounds on the $\d$'s,   at fixed values
of $M_{sq}$ and $m_{\tilde g}$, with the condition $M_{sq} > m_{\tilde g}$.
In this way one gets a set of  constraints on individual $\d$'s. 
Indeed, since we are interested in model independent
constraints,  it is meaningful to study the interference
of  cancellation effects  only in specific models.

The physical condition used to get
the bounds on the $\delta$'s is  that the SUSY contribution (proportional to  each single $\d$)
 plus the SM
contribution to $\Delta M_K$ and $\eps_K$ do not exceed the experimental value
of these quantities.
For what concerns the SM contribution to $\Delta M_K$, we assume that the values of the
CKM elements $V_{cd}$ and $V_{cs}$ are unaffected by SUSY.
This implies the (very reasonable) hypothesis that SUSY does not correct significantly
tree level weak decays.
The value of the SM contribution to $\eps_K$, instead, depends on the
phase of the CKM matrix. 
This phase can be largely affected by unknown SUSY corrections and can be treated as a
free parameter. We  put the CKM phase to zero so that 
the experimental value of $\eps_K$ is completely determined by SUSY.
Finally, to be even more  conservative, we subtract one
standard deviation to the values of the B-parameters.

The final results are shown in tabs.~\ref{tab:m250}-\ref{tab:e1000} for gluino
masses of 250, 500, 1000 GeV.
We consider the  heavy squark masses expected in some common  models
(see {\it e.g.}~\cite{dine,dimo,poma,cohe}).

The constraints that come from the four possible insertions of the $\d$'s are presented:
in the first and second rows only terms proportional respectively to $\ded_{LL}$ and
$\ded_{LR}$ are considered; in the last two rows the contribution of  operators with
opposite chirality, $RR$ and $RL$, is also evaluated by assuming
$\ded_{LR}=\ded_{RL}$ and $  \ded_{LL}=\ded_{RR}$.

In each column of the table we show the bounds on the $\d$'s in the various approximations
that one can use for their determination: 
without QCD correction and in VIA,
 with LO-QCD corrections and in VIA, with LO-QCD corrections
and with lattice B-parameters and, eventually, with NLO-QCD corrections and lattice B-parameters.
Comparing the values of our constraints at LO-VIA  with those found from
the authors of ref.~\cite{bagg} we find some differences.
The reason is twofold.
On the one hand  they do not consider  the SM contribution to $\Delta M_K$ and 
on the other  they evaluate the hadronic matrix elements at a scale $\tilde{\mu}$
such that $\a_s(\tilde{\mu})=1$. This latter choice may be questionable
because at this scale strong interactions break perturbation theory.

The combination of B-parameters and NLO-QCD corrections change the LO-VIA results by 
about $25-35\%$. As expected~\cite{bagg}, the tightest constraints are for
the cases $\ded_{LL}=\ded_{RR}$ and $\ded_{LR}=\ded_{RL}$.
In these cases the coefficients proportional to $\ded_{LL}\ded_{RR}$,
$\ded_{LR}\ded_{RL}$ dominate the others.

We have checked that the uncertainties of the results due to higher perturbative orders,
are sizeable, being, in some cases up to $10\%$.

\section{Constraints on squarks spectrum}

\label{sec:spec}
In this section, following the discussion of
ref.~\cite{arka}, we provide a different kind of constraints.

For fixed values  of the $\d$'s and of the average light sparticle mass,
$m_{\tilde g}$, it is possible to calculate the minimum value of $M_{sq}$ necessary to
 suppress the FCNC at an experimentally acceptable level.  
 Here we give constraints on $M_{sq}$ and we discuss about their consistency. 
Using Renormalization Group Equations, one finds that
a too large $M_{sq}$  can drive  to zero or negative values the average mass of 
the third generation of sfermions, $m_{\tilde f}$, 
at the TeV scale ($m_{\tilde f}(\sim 1 {\rm{TeV}})$).
To circumvent this  problem, a minimum value for $m_{\tilde f}(\mu_{\rm{GUT}})$
at the GUT scale  has to  be chosen.
If $m_{\tilde f}(\mu_{\rm{GUT}})$ is too high (say more then 3-4 TeV), however, a too
large fine-tuning  of the SUSY parameters is required in order to account
for the observed mass of the $Z$-boson and severe naturalness problems
arise~\cite{barb,ande}. This problem was studied in refs.~\cite{arka,agas}. 

One obtains constraints about the consistency of models 
with a splitted mass spectrum  following  three steps:
\begin{itemize}
\item
determining  the {\it minimum} value of $M_{sq}$ necessary
to suppress FCNC. This is discussed in subsec.~\ref{sub:min};
\item
computing the {\it maximum} value of $M_{sq}$ allowed by positiveness of light scalar 
masses and fine-tuning. More about this in subsec.~\ref{sub:rge};
\item
combining the previous two results one can determine
regions of allowable values of $M_{sq}$ that satisfy both the requests of the
previous points. We comment about that  in subsec.~\ref{sub:res}. 
\end{itemize}

\subsection{Minimum values for heavy squark mass}

\label{sub:min}

\begin{figure}[t]
\begin{center}
\hspace*{-2.5cm}
\begin{minipage}[t]{0.12\linewidth}
\vspace*{-5.5cm}
$M_{sq} $~[TeV]
\end{minipage}
\hspace*{-0.5cm}
\epsfig{file=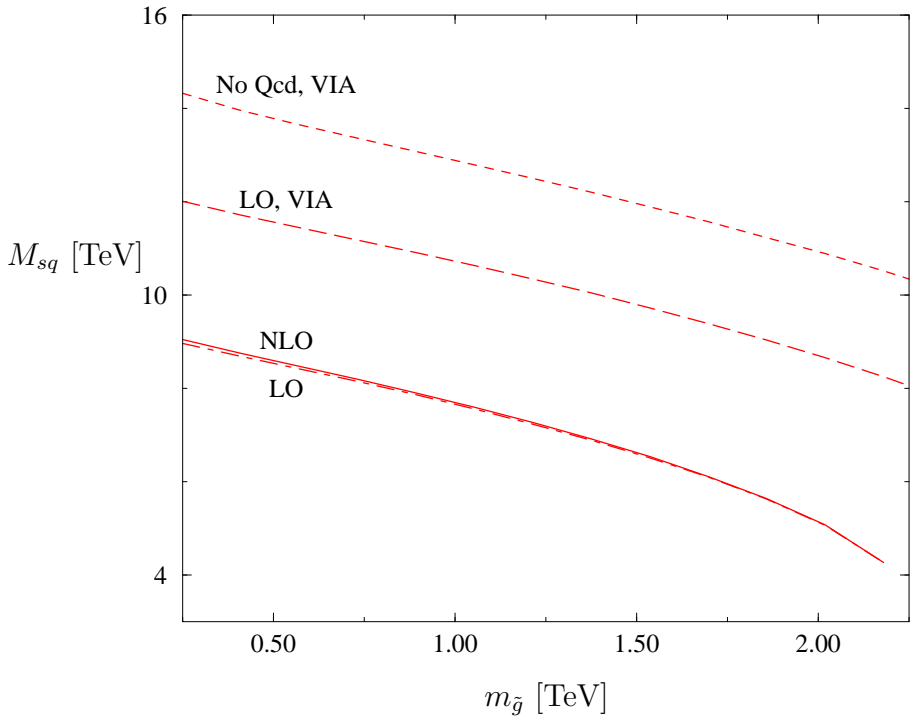,width=0.62\linewidth} \\[5pt]
$\msg$ [TeV]
\caption[]{\it  Lower bounds  on $M_{sq}$  from $\Delta M_K$ 
with various approximations
for the case I with ${\cal{K}}=0.22$. 
In this case the larger correction to
LO-VIA come from the B-parameters.}
\protect\label{fig:confr1}
\end{center}
\end{figure}
\begin{figure}[t]
\begin{center}
\hspace*{-2.5cm}
\begin{minipage}[t]{0.12\linewidth}
\vspace*{-5.5cm}
$M_{sq} $~[TeV]
\end{minipage}
\hspace*{-0.5cm}
\epsfig{file=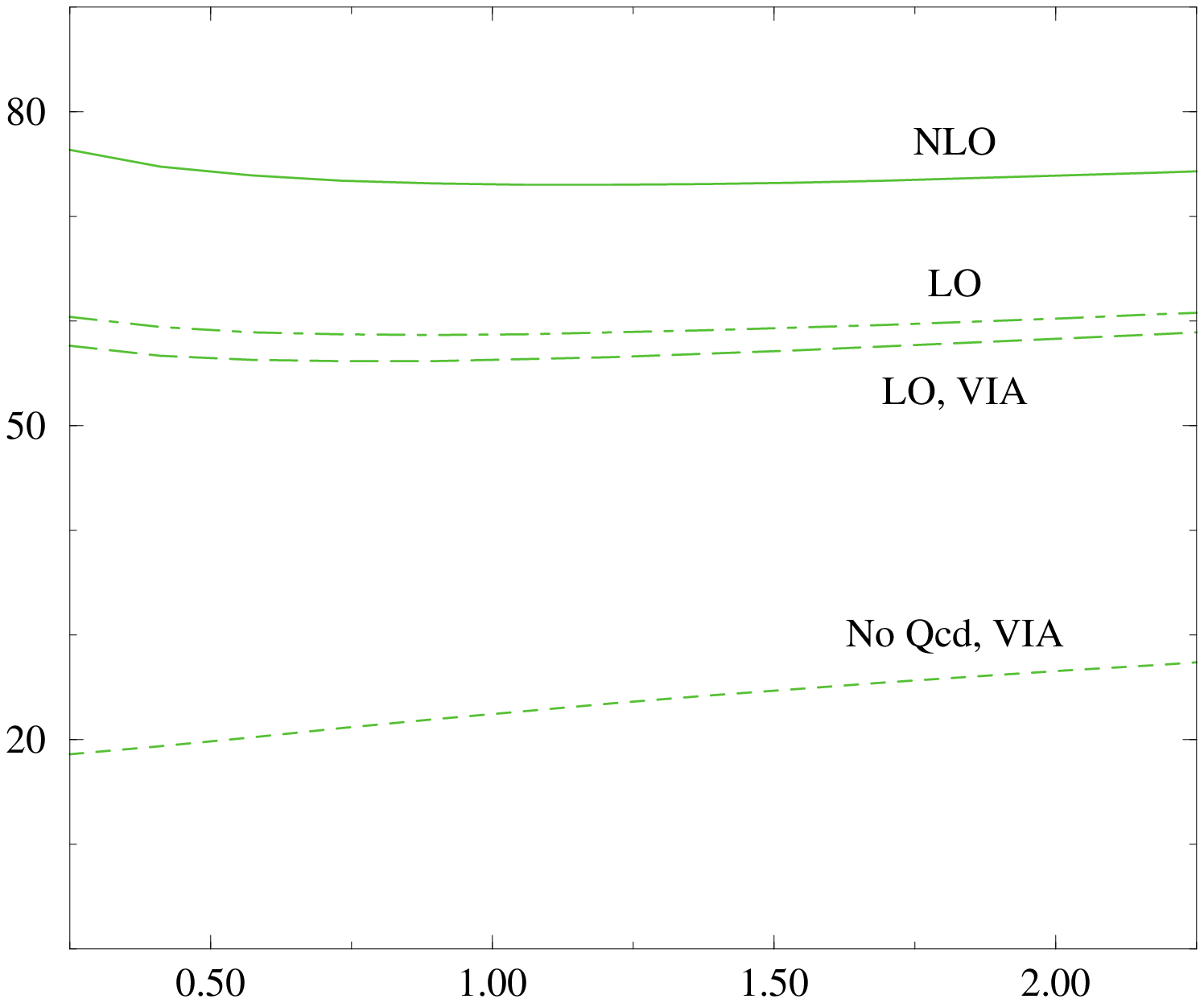,width=0.62\linewidth} \\[5pt]
$\msg$ [TeV]
\caption[]{\it  Lower bounds on $M_{sq}$  from $\Delta M_K$ 
with various approximations
for the case III with ${\cal{K}}=0.22$.
In this case the larger correction
to LO-VIA come from NLO perturbative corrections.}
\protect\label{fig:confr3}
\end{center}
\end{figure}
In order to obtain constraints on $M_{sq}$ one has
 to specify a  value for the $\d$'s.
We consider the  cases  
\begin{equation}
\mbox{
\begin{tabular}{cccccc}
	& $\ded_{LL}$& $\ded_{LR}$& $\ded_{RL}$& $\ded_{RR}$ \\[.4cm] 
I	& $ {\cal{K}} $& 0& 0& 0\\
II	& 0&${\cal{K}}$ & 0& 0&\\
III	& ${\cal{K}}$ & 0& 0& ${\cal{K}}$\\
IV	& 0&${\cal{K}}$&${\cal{K}}$&0\\
\end{tabular}
}
\label{cases}
\end{equation}
where ${\cal{K}}$ can take the values (1, 0.22, 0.05). We have chosen these entries
to leave aside  possible accidental cancellations.
The cases in which  $\cal{K}$=1, are, of course,  extreme cases:
one may wonder about the consistency  of mass insertion approximation
as the neglected terms are of  order ${\cal{O}}(1)$.
However these cases have already been studied in the literature, (see {\it e.g.}
\cite{arka,agas}),  and we report them for
completeness. 
The results so obtained just give an estimate of the mass scales that are involved and can be
trusted if other corrections do not provide accidental cancellations.  
This can be checked only by a direct calculation.  

The assumptions made for   the SM contribution and the  B-parameters are the same as in 
section~\ref{sec:deltas}.

\begin{figure}[t]
\begin{center}
\hspace*{-2.5cm}
\begin{minipage}[t]{0.12\linewidth}
\vspace*{-5.5cm}
$M_{sq} $~[TeV]
\end{minipage}
\hspace*{-0.5cm}
\epsfig{file=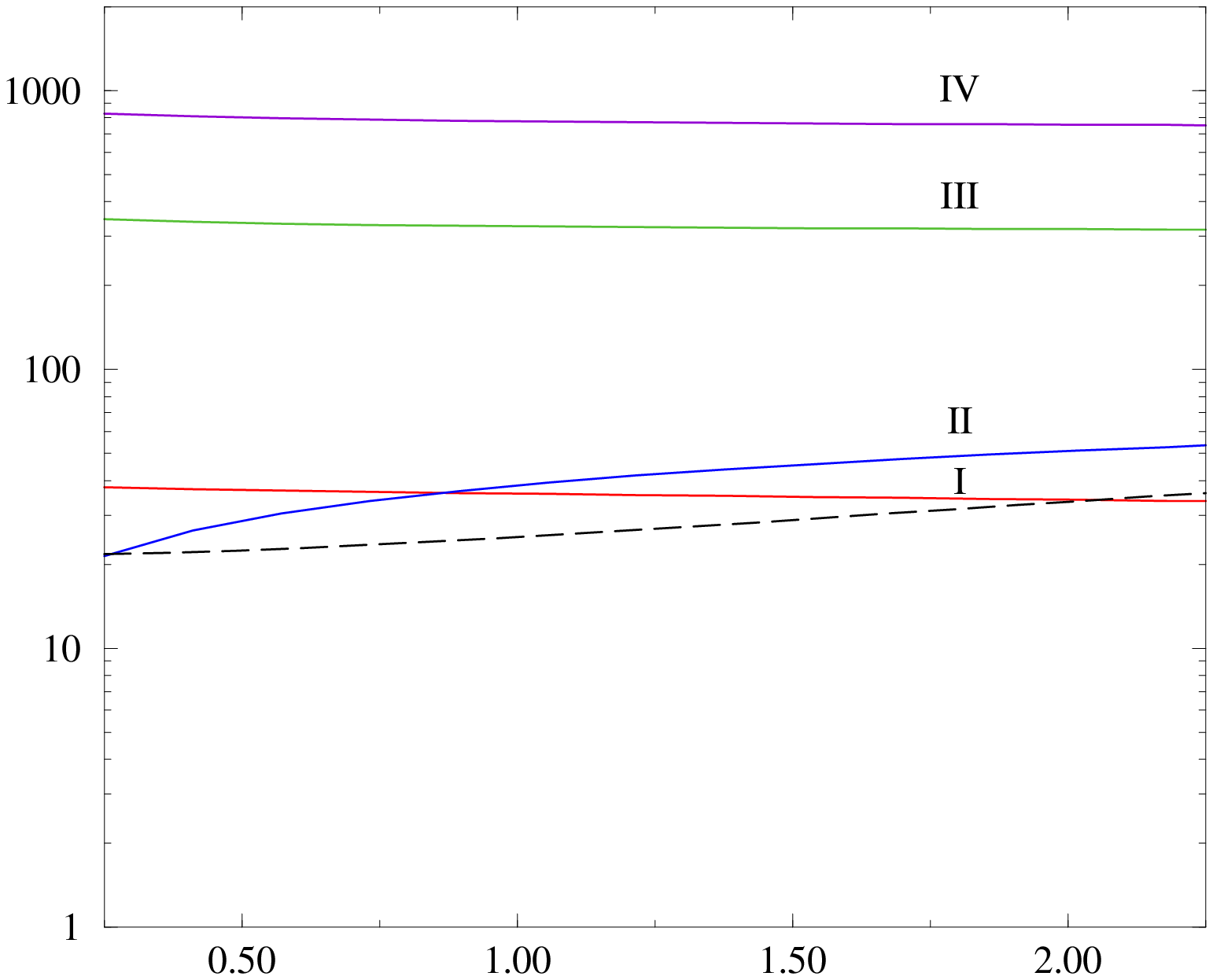,width=0.60\linewidth} \\[5pt]
$\msg$ [TeV]
\caption[]{\it The full (colored) lines give the lower bounds on $M_{sq}$ necessary to suppress FCNC
and with ${\cal{K}}$=1  for the 
various cases.
 An upper bound on  $M_{sq}$ is derived  
 in order to  satisfy fine-tuning requirements and it is shown by the dashed line.
 The two kind on constraints are not compatible in this case.}
\protect\label{fig:caso1}
\end{center}
\end{figure}
\begin{figure}
\begin{center}
\hspace*{-2.5cm}
\begin{minipage}[t]{0.12\linewidth}
\vspace*{-5.5cm}
$M_{sq} $~[TeV]
\end{minipage}
\hspace*{-0.5cm}
\epsfig{file=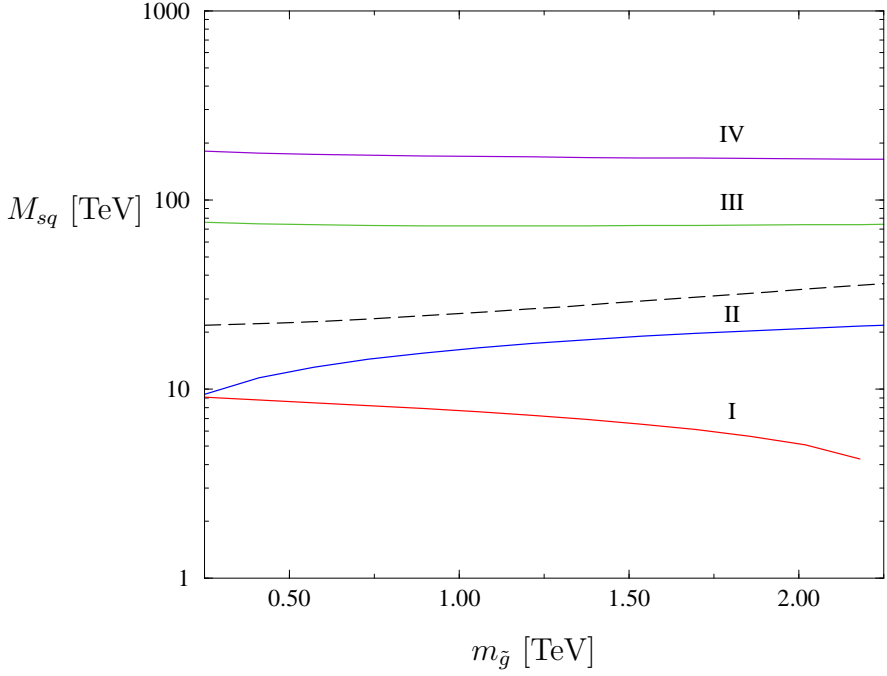,width=0.60\linewidth} \\[5pt]
$\msg$ [TeV]
\caption[]{\it  The same as fig.~\ref{fig:caso1} with ${\cal{K}}$=0.22.
Cases I and II are now compatible with fine-tuning requirements. }
\protect\label{fig:caso022}
\end{center}
\end{figure}
\begin{figure}
\begin{center}
\hspace*{-2.5cm}
\begin{minipage}[t]{0.12\linewidth}
\vspace*{-5.5cm}
$M_{sq} $~[TeV]
\end{minipage}
\hspace*{-0.5cm}
\epsfig{file=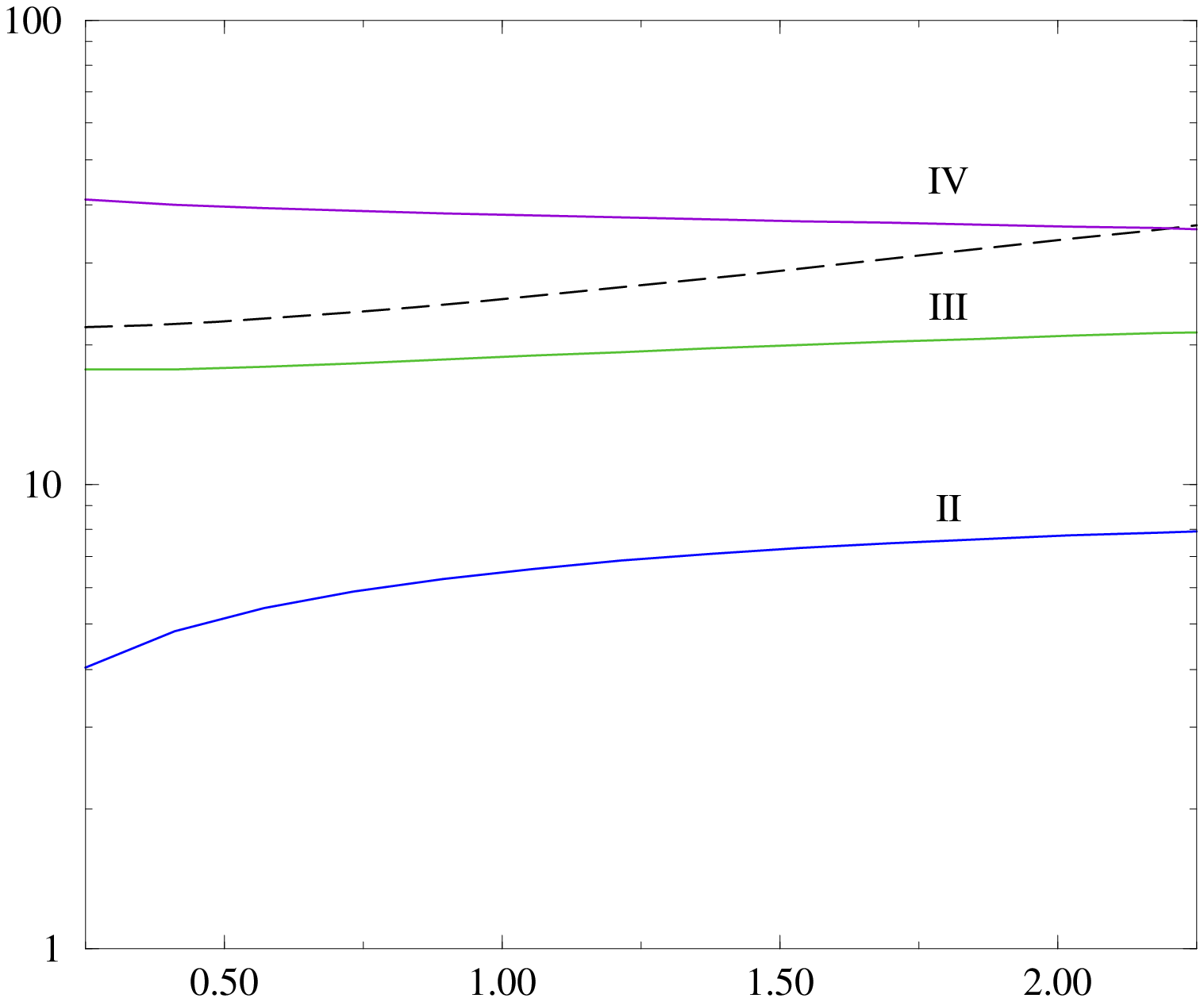,width=0.60\linewidth} \\[5pt]
$\msg$ [TeV]
\caption[]{\it The same as fig.~\ref{fig:caso1} with ${\cal{K}}$=0.05.
Case I is not drawn since no lower bound on $M_{sq}$ 
can be obtained in this case.
 Cases II and III are now compatible with fine-tuning requirements. }
\protect\label{fig:caso005}
\end{center}
\end{figure}

In order to monitor the effect of the different   corrections on the final result 
 we show in figs.~\ref{fig:confr1},~\ref{fig:confr3} the
lower bound obtained  for the cases I and III with ${\cal{K}}=0.22$ (the other cases give
similar results).
As we see, B-parameters and NLO-QCD corrections play  a significant
r\^ole in the final computation and the correction they provide with respect to
the LO-VIA results are of the order of $(25-35) \%$. 
In particular, in case I, fig.~\ref{fig:confr1}, B-parameters provide
the most important corrections with respect to LO-VIA results.
In case III, fig.~\ref{fig:confr3}, instead, 
corrections to LO-VIA results are dominated by the NLO-QCD perturbative
contributions. 

The  case $\ded_{LL}=\ded_{RR}$ was also  considered in refs.~\cite{arka,agas}.
The  differences, at LO and without B-parameters, among our result and the ones of
refs.~\cite{arka,agas} come from our inclusion of  the SM contribution, from
the value of the strange quark mass and from the scale at which hadronic matrix elements 
are evaluated. We agree with them for the same choice of parameters.

Notice that, if the imaginary parts of the $\d$'s are of the same order of their real
parts, there are much stronger constraints coming from $\eps_K$ than from
$\Delta M_K$ (namely by a factor
$\sim 7.7$). To be conservative, we consider in this section  only constraints coming
from the real parts of the $\d$'s.

The final results are shown by the (colored) continuous lines in
figs.~\ref{fig:caso1},~\ref{fig:caso022},~\ref{fig:caso005}.  
The minimum value of $M_{sq}$ depends strongly  both on ${\cal{K}}$ and
on the case one considers (I, II, III or IV, see eq.~\ref{cases}).
Notice that $\ded_{LR}$, $\ded_{RL}$, (which enter the cases II and IV), are ``naturally''
small in the MSSM.
However, since we would like to do a model--independent analysis, we have made no
particular assumption on them.
In all  graphs the strongest constraints come from the case 
$\ded_{LR}=\ded_{RL}\neq 0$. 
Much lower constraints are generally obtained in cases I and II.
In fig.~\ref{fig:caso005} the case I has not been drawn 
since no constraint can be derived.

\subsection{RGE for the masses of the third generation of scalars}
 
\label{sub:rge}
It is well known that large values of $M_{sq}$ can drive the mass of the third
generation of scalars to negative values, via RGE~\cite{arka}. 
Let us  consider the two-loop RGE's  for the mass, $m_{\tilde f}$ of the
third generation of scalars, $\tilde{f}$. In the $\overline{DR}^{\ \prime}$
scheme (see {\it e.g.} ref.~\cite{sieg}), with two generations of heavy scalars, one has 
\bea
\mu \fr{ d}{d \mu}m_{\tilde f}(\mu) &=&
-\fr{8}{4 \pi} \sum_i \a_i(\mu)\  C_i^{\tilde f} 
(m_{G}^2)_i(\mu)+
\fr{32}{(4 \pi)^2} \sum_i\a_i^2(\mu)\  C_i^{\tilde f} M_{sq}^2,
\label{eq:rge}
\eea
where $ C_i^{\tilde f}$ is the Casimir factor for $\tilde f$ in the SU(5) normalization, the
sums are over the gauge groups SU(3), SU(2), U(1) and $m_G$ denotes the gaugino masses.
In eq.~(\ref{eq:rge}), Yukawa couplings are neglected: these couplings drive the light
masses to even lower values and so, in this respect, our choice  is a conservative one.
Moreover, the introduction  of Yukawa interactions requires 
further assumptions on SUSY parameters (see {\it e.g.}~\cite{agas})
that we  do not  discuss in this paper. 

The solution of eq.~(\ref{eq:rge}) between a Grand Unification (GUT) scale 
$\mu_{\rm GUT}\sim 2 \cdot 10^{16}$ GeV, and $\mu\sim 1$ TeV can be easily written as
\bea
m_{\tilde f}^2(\mu)&=&m_{\tilde f}^2(\mu_{\rm{GUT}})-
 \sum_i \fr{16}{ 4 \pi \b_i^0} \Big[\a_i(\mu_{\rm{GUT}})-\a_i (M_{sq})\Big] 
C_i^{\tilde f} \ M_{sq}^2 +\nn \\
& & \sum_i \fr{2}{\b^0_i} \Big[ m_{G}^2(\mu_{\rm{GUT}})-(m_{G}^2)_i(M_{sq})\Big]
C_i^{\tilde f} + \nn \\
& & \sum_i \fr{2}{\b_i^0} \Big[ (m_{G})_i^2(M_{sq})-(m_{G}^2)_i(\mu)\Big] C_i^{\tilde f},
\label{eq:rgefin}
\eea
where $\b_i^0$ are the $\b$-functions LO coefficients of the $i$-th gauge coupling.
In eq.~(\ref{eq:rgefin}) we have considered a common gaugino mass, $m_G$ at the GUT
scale, while for what concerns the couplings we have evolved them starting backward  
from $\mu=M_Z$. Note that in eq.~(\ref{eq:rgefin}) the contribution of the heavy scalars 
has been decoupled  at  $ M_{sq}$.

Eq.~(\ref{eq:rgefin}) can be used in order to derive consistency 
constraints on the values of $M_{sq}$ and $m_{G}^2(\mu_{\rm{GUT}})$ once the 
values of $m_{\tilde f}^2(\mu)$ and 
of $m_{\tilde f}^2(\mu_{\rm{GUT}})$ are fixed.
The latter can be determined according to the following requirements. 
First, $m_{\tilde f}^2(\mu)$ must be at least positive, such as to leave color and
electric symmetries unbroken.
The value of $m_{\tilde f}^2(\mu_{\rm{GUT}})$  determines the amount of  fine-tuning
necessary in order to achieve  the electroweak symmetry breaking.
Following ref.~\cite{barb} the necessary fine-tuning scales approximately as
$10\% \times (0.3 \ {\rm {TeV}}/m_{\tilde{Q}_3}(\mu_{\rm{GUT}}))^2 $ for 
the squark doublet of the third generation $\tilde{Q}_3$.
We have calculated  the constraints on $M_{sq}$ and $m_{G}^2(\mu_{\rm{GUT}})$ coming 
from eq.~(\ref{eq:rgefin}) in the case ${\tilde{f}}={\tilde{Q}_3 }$ 
choosing for $m_{\tilde{Q}_3 }^2(\mu_{\rm{GUT}})$ the value of
$(3.5 \ {\rm {TeV}})^2$. The latter choice corresponds to a fine-tuning of more than
$0.1\%$.

At fixed values of $m_{\tilde{Q}_3 }^2(\mu)$ and $m_{\tilde{Q}_3 }^2(\mu_{\rm{GUT}})$
(which depend on $M_{sq}$ and $m_{G}$) one can plot the upper value of $M_{sq}$ as
function of $m_{G}$. The result is the   (black) dashed line of 
figs.~\ref{fig:caso1},~\ref{fig:caso022},~\ref{fig:caso005}.
One finds that $M_{sq}$  can not be much larger  than about 25 TeV.
Of course this is just an estimate of this limiting value.
The inclusion of Yukawa couplings, of more severe fine-tuning requirements 
and of other effects can only lower this limit.

\subsection{Final  remarks}

\label{sub:res}
In figs.~\ref{fig:caso1},~\ref{fig:caso022},~\ref{fig:caso005}
we combine the constraints derived in the two previous subsections.
These figures (together with tables tabs.~\ref{tab:m250}-\ref{tab:e1000}) suggest
that also models with a splitted mass spectrum need further assumptions
to be phenomenologically viable, {\it e.g.} one has to introduce
flavor symmetry or dynamical generation of degenerate scalar masses~\cite{arka}.

In particular, without these further hypotheses, most of the cases 
which we have considered, face fine tuning problems.
In particular, values of ${\cal{K}}\sim {\cal{O}}(1)$  are hardly acceptable.
Although ${\cal{K}}\sim {\cal{O}}(0.22)$ and ${\cal{K}}\sim {\cal{O}}(0.05)$  have better
chances they must be treated carefully.

\section{Conclusions}

\label{sec:conc}
In this work we  analyze in detail the constraints on SUSY-models
parameters coming from $\rm K-\overline{K}$ oscillations in the hypothesis
of a splitted SUSY spectrum.  FCNC contributions coming from
gluino-squark-quark interactions, working in the so-called mass insertion approximation,
have been considered. We  provide boundaries on mass insertions  and
on SUSY mass scales and we  discus their consistency.
Previous results including NLO-QCD corrections to $\Delta S=2$ effective Hamiltonian,
B-parameters for the evaluation of hadronic matrix elements have been improved.
A full set of magic-numbers is provided, that can be used for further analyses.

We have discussed the residual uncertainty of our results
coming from our ignorance of NLO-QCD corrections to the matching coefficients.

Our analysis confirms that  a splitted sparticle mass spectrum
does not explain easily  FCNC suppression without some amount of  fine-tuning.
These problems can be solved only if further 
assumptions in these kind of models are made, {\it e.g.}
flavor symmetry or dynamical generation of degenerate scalar masses~\cite{arka}. 

In order  to perform a complete analysis of SUSY-FCNC effects 
chargino contributions should be included. It is also interesting 
to extend  this kind of analysis to $\Delta {\rm B}=2$ processes, once 
the  calculation of B-parameters for the $B-\overline{B}$
system parameters (which is in progress~\cite{reti}) will be completed. 

\subsection*{Acknowledgements}
R.~C. wants to thank INFN for partial support.
M.~Ciuchini, E.~Franco, V.~Lubicz,
G.~Martinelli, A.~Masiero and L.~Silvestrini are acknowledged for  fruitful discussions.

\appendix
\section*{Appendix}

We give here the numerical values for the magic numbers of eq.~(\ref{eq:magic}).
Only the non-vanishing entries are shown: 
\begin{displaymath}
\begin{array}{rclrcl}
a_{(r)} 	&=& (0.29, -1.1, 0.14, -0.69, 0.79) &&&\\
a_{(r)}^\prime 	&=& (0.46, -1.8, 0.23, -1.1, 1.3)   &&&\\[0.8cm]
b_{11}^{(r)} &=& (1.5, 0, 0, 0, 0) &
c_{11}^{(r)} &=& (-3.4, 0, 0, 0, 0) \\
b_{22}^{(r)} &=& (0, 0.0048, 1.1, 0, 0) &
c_{22}^{(r)} &=& (0, -0.12, 2.8, 0, 0) \\
b_{23}^{(r)} &=& (0, -0.0073, 0, 0, 0) &
c_{23}^{(r)} &=& (0, 0.12, 1.2, 0, 0) \\
b_{32}^{(r)} &=& (0, -0.23, 0.47, 0, 0) &
c_{32}^{(r)} &=& (0, 6.3, 2.0, 0, 0) \\
b_{33}^{(r)} &=& (0, 0.34, 0, 0, 0) &
c_{33}^{(r)} &=& (0, -6.2, 0.88, 0, 0) \\
b_{44}^{(r)} &=& (0, 0, 0, 0.52, -0.017) &
c_{44}^{(r)} &=& (0, 0, 0, -5.7, 1.8) \\
b_{45}^{(r)} &=& (0, 0, 0, 0.99, -2.2) &
c_{45}^{(r)} &=& (0, 0, 0, -26, -27) \\
b_{54}^{(r)} &=& (0, 0, 0, -0.00051, 0.020) &
c_{54}^{(r)} &=& (0, 0, 0, 0.0086, -0.77) \\
b_{55}^{(r)} &=& (0, 0, 0, -0.00096, 2.5) &
b_{55}^{(r)} &=& (0, 0, 0, 0.040, 12) \\[0.8cm]
d_{11}^{(r)} &=& (1.0, 0, 0, 0, 0) &
e_{11}^{(r)} &=& (-1.5, 0, 0, 0, 0) \\
d_{22}^{(r)} &=& (0, 0, 1.0, 0, 0) &
e_{22}^{(r)} &=& (0, 0.70, -4.9, 0, 0) \\
d_{23}^{(r)} &=& (0, 0, 0, 0, 0) &
e_{23}^{(r)} &=& (0, -1.1, 0, 0, 0) \\
d_{32}^{(r)} &=& (0, -0.67, 0.67, 0, 0) &
e_{32}^{(r)} &=& (0, -27, -9.0, 0, 0) \\
d_{33}^{(r)} &=& (0, 1.0, 0, 0, 0) &
e_{33}^{(r)} &=& (0, 40, 0, 0, 0) \\
d_{44}^{(r)} &=& (0, 0, 0, 1.0, -0.015) &
e_{44}^{(r)} &=& (0, 0, 0, 22, 0.46) \\
d_{45}^{(r)} &=& (0, 0, 0, 1.9, -1.9) &
e_{45}^{(r)} &=& (0, 0, 0, 41, 58) \\
d_{54}^{(r)} &=& (0, 0, 0, -0.0081, 0.0081) &
e_{54}^{(r)} &=& (0, 0, 0, 0.15, -0.12) \\
d_{55}^{(r)} &=& (0, 0, 0, -0.015, 1.0) &
e_{55}^{(r)} &=& (0, 0, 0, 0.29, -15) \\[0.8cm]
f_{11}^{(r)} &=& (1.5, 0, 0, 0, 0) &&& \\
f_{22}^{(r)} &=& (0, 0, 4.2, 0, 0) &&& \\
f_{23}^{(r)} &=& (0, 0, 1.1, 0, 0) &&& \\
f_{32}^{(r)} &=& (0, 33, 2.8, 0, 0) &&& \\
f_{33}^{(r)} &=& (0, -41, 0.70, 0, 0) && \\
f_{44}^{(r)} &=& (0, 0, 0, -23, 0.40) &&& \\
f_{45}^{(r)} &=& (0, 0, 0, -72, -28) &&& \\
f_{54}^{(r)} &=& (0, 0, 0, 0.18, -0.21) &&& \\
f_{55}^{(r)} &=& (0, 0, 0, 0.57, 15) &&& 
\end{array}
\end{displaymath}

{\small
\begin{table} 
\begin{center} 
\begin{tabular}{|c|c|c|c|c|} 
\hline 
$M_{sq}$ [TeV] & No-QCD, VIA & LO-VIA & LO,     $B_i$ & NLO, $B_i$ \\ 
\hline \hline\multicolumn{5}{|c|}{$\sqrt{|{\rm{Re}} (\delta^d_{12})^2_{LL}|}$} \\ 
\hline 
2 & $3.1\times 10^{-2}$ & $3.6\times 10^{-2}$ & $4.9\times 10^{-2}$ & $4.9\times 10^{-2}$ \\ 
 5 & $7.5\times 10^{-2}$ & $8.8\times 10^{-2}$ & 0.12 & 0.12\\ 
 10 & 0.15 & 0.18&0.25 & 0.24\\ 
\hline \hline 
\multicolumn{5}{|c|}{$\sqrt{|{\rm{Re}} (\delta^d_{12})^2_{LR}|}$} \\ 
\hline 
2 & $2.1\times 10^{-2}$ & $1.4\times 10^{-2}$ & $1.8\times 10^{-2}$ & $1.6\times 10^{-2}$ \\ 
 5 & $9.8\times 10^{-2}$ & $6.5\times 10^{-2}$ & $8.2\times 10^{-2}$ & $7.2\times 10^{-2}$\\ 
 10 & 0.34 & 0.22&0.28 & 0.25\\ 
\hline \hline 
\multicolumn{5}{|c|}{$\sqrt{|{\rm{Re}} (\delta^d_{12})_{LR}={\rm{Re}} (\delta^d_{12})_{RL}|}$} \\ 
\hline
 2 & $6.6\times 10^{-3}$ & $3.5\times 10^{-3}$ & $3.5\times 10^{-3}$ & $2.8\times 10^{-3}$ \\ 
 5 & $1.5\times 10^{-2}$ & $7.7\times 10^{-3}$ & $7.9\times 10^{-3}$ & $6.4\times 10^{-3}$\\ 
 10 & $3.0\times 10^{-2}$ & $1.5\times 10^{-2}$&$1.5\times 10^{-2}$ & $1.2\times 10^{-2}$\\ 
\hline \hline 
\multicolumn{5}{|c|}{$\sqrt{|{\rm{Re}} (\delta^d_{12})^2_{LL}={\rm{Re}} (\delta^d_{12})^2_{RR}|}$} \\ 
\hline 
2 & $1.1\times 10^{-2}$ & $5.2\times 10^{-3}$ & $5.1\times 10^{-3}$ & $4.1\times 10^{-3}$ \\ 
 5 & $4.1\times 10^{-2}$ & $1.6\times 10^{-2}$ & $1.6\times 10^{-2}$ & $1.3\times 10^{-2}$\\ 
 10 & 0.10 & $3.6\times 10^{-2}$&$3.4\times 10^{-2}$ & $2.7\times 10^{-2}$\\ 
\hline \hline 
\end{tabular} 
\caption{{\it Limits on ${\rm Re }  (\delta^d_{12})_{AB}$ from $\Delta M_K$ with gaugino masses of 250 
  GeV.}}
\label{tab:m250}
\end{center} 
\end{table} 
\begin{table} 
\begin{center} 
\begin{tabular}{|c|c|c|c|c|} 
\hline 
$M_{sq}$ [TeV] & No-QCD, VIA & LO-VIA & LO,     $B_i$ & NLO, $B_i$ \\ 
\hline \hline\multicolumn{5}{|c|}{$\sqrt{|{\rm{Re}} (\delta^d_{12})^2_{LL}|}$} \\
 \hline 
 2 & $3.8\times 10^{-2}$ & $4.5\times 10^{-2}$ & $6.1\times 10^{-2}$ & $6.1\times 10^{-2}$ \\ 
 5 & $8.1\times 10^{-2}$ & $9.6\times 10^{-2}$ & 0.13 & 0.13\\ 
 10 & 0.16 & 0.19&0.26 & 0.26\\ 
\hline \hline 
\multicolumn{5}{|c|}{$\sqrt{|{\rm{Re}} (\delta^d_{12})^2_{LR}|}$} \\ 
\hline
 2 & $1.6\times 10^{-2}$ & $1.1\times 10^{-2}$ & $1.3\times 10^{-2}$ & $1.2\times 10^{-2}$ \\ 
 5 & $6.3\times 10^{-2}$ & $4.2\times 10^{-2}$ & $5.3\times 10^{-2}$ & $4.7\times 10^{-2}$\\ 
 10 & 0.21 & 0.14&0.17 & 0.15\\ 
\hline \hline 
\multicolumn{5}{|c|}{$\sqrt{|{\rm{Re}} (\delta^d_{12})_{LR}={\rm{Re}} (\delta^d_{12})_{RL}|}$} \\ 
\hline 
2 & $9.6\times 10^{-3}$ & $4.6\times 10^{-3}$ & $4.5\times 10^{-3}$ & $3.6\times 10^{-3}$ \\ 
 5 & $1.7\times 10^{-2}$ & $8.5\times 10^{-3}$ & $8.7\times 10^{-3}$ & $7.0\times 10^{-3}$\\ 
 10 & $3.2\times 10^{-2}$ & $1.6\times 10^{-2}$&$1.6\times 10^{-2}$ & $1.3\times 10^{-2}$\\ 
\hline \hline 
\multicolumn{5}{|c|}{$\sqrt{|{\rm{Re}} (\delta^d_{12})^2_{LL}={\rm{Re}} (\delta^d_{12})^2_{RR}|}$} \\ 
\hline 
2 & $8.6\times 10^{-3}$ & $4.4\times 10^{-3}$ & $4.4\times 10^{-3}$ & $3.6\times 10^{-3}$ \\ 
 5 & $3.2\times 10^{-2}$ & $1.4\times 10^{-2}$ & $1.4\times 10^{-2}$ & $1.1\times 10^{-2}$\\ 
 10 & $8.8\times 10^{-2}$ & $3.4\times 10^{-2}$&$3.2\times 10^{-2}$ & $2.6\times 10^{-2}$\\ 
\hline \hline 
\end{tabular} 
\caption{{\it Limits on ${\rm Re }  (\delta^d_{12})_{AB}$ from $\Delta M_K$ with gaugino
 masses of 500 GeV.}}
\label{tab:m500}
\end{center} 
\end{table} 
\begin{table} 
\begin{center} 
\begin{tabular}{|c|c|c|c|c|} 
\hline 
$M_{sq}$ [TeV] & No-QCD, VIA & LO-VIA & LO,     $B_i$ & NLO, $B_i$ \\ 
\hline \hline\multicolumn{5}{|c|}{$\sqrt{|{\rm{Re}} (\delta^d_{12})^2_{LL}|}$} \\ 
\hline
 2 & $5.9\times 10^{-2}$ & $6.9\times 10^{-2}$ & $9.4\times 10^{-2}$ & $9.3\times 10^{-2}$ \\ 
 5 & $9.6\times 10^{-2}$ & 0.11 & 0.15 & $0.15$\\ 
 10 & 0.17 & 0.21&0.28 & $0.28$\\ 
\hline \hline 
\multicolumn{5}{|c|}{$\sqrt{|{\rm{Re}} (\delta^d_{12})^2_{LR}|}$} \\ 
\hline 
2 & $1.4\times 10^{-2}$ & $9.7\times 10^{-3}$ & $1.2\times 10^{-2}$ & $1.1\times 10^{-2}$ \\ 
 5 & $4.6\times 10^{-2}$ & $3.0\times 10^{-2}$ & $3.9\times 10^{-2}$ & $3.4\times 10^{-2}$\\ 
 10 & 0.14 & $8.8\times 10^{-2}$&0.11 & $9.8\times 10^{-2}$\\ 
\hline \hline 
\multicolumn{5}{|c|}{$\sqrt{|{\rm{Re}} (\delta^d_{12})_{LR}={\rm{Re}} (\delta^d_{12})_{RL}|}$} \\ 
\hline
 2 & $4.2\times 10^{-2}$ & $9.8\times 10^{-3}$ & $7.8\times 10^{-3}$ &$6.0\times 10^{-3}$ \\ 
 5 & $2.2\times 10^{-2}$ & $1.1\times 10^{-2}$ & $1.1\times 10^{-2}$ & $8.5\times 10^{-3}$\\ 
 10 & $3.6\times 10^{-2}$ & $1.8\times 10^{-2}$&$1.8\times 10^{-2}$ & $1.4\times 10^{-2}$\\ 
\hline \hline 
\multicolumn{5}{|c|}{$\sqrt{|{\rm{Re}} (\delta^d_{12})^2_{LL}={\rm{Re}} (\delta^d_{12})^2_{RR}|}$} \\ 
\hline
 2 & $8.0\times 10^{-3}$ & $4.2\times 10^{-3}$ & $4.3\times 10^{-3}$ & $3.5\times 10^{-3}$ \\ 
 5 & $2.5\times 10^{-2}$ & $1.2\times 10^{-2}$ & $1.2\times 10^{-2}$ & $9.7\times 10^{-3}$\\ 
 10 & $6.8\times 10^{-2}$ & $2.9\times 10^{-2}$&$2.9\times 10^{-2}$ & $2.3\times 10^{-3}$\\ 
\hline \hline 
\end{tabular} 
\caption{{\it Limits on ${\rm Re }  (\delta^d_{12})_{AB}$ from $\Delta M_K$ with gaugino masses of 1000   GeV.}}
\label{tab:m1000}
\end{center} 
\end{table}

\begin{table} 
\begin{center} 
\begin{tabular}{|c|c|c|c|c|} 
\hline 
$M_{sq}$ [TeV] & No-QCD, VIA & LO-VIA & LO,     $B_i$ & NLO, $B_i$ \\ \hline \hline
\multicolumn{5}{|c|}{$\sqrt{|{\rm{Im}} (\delta^d_{12})^2_{LL}|}$} \\ 
\hline 
2 & $4.0\times 10^{-3}$ & $4.7\times 10^{-3}$ & $6.4\times 10^{-3}$ & $6.4\times 10^{-3}$ \\ 
 5 & $9.7\times 10^{-3}$ & $1.2\times 10^{-2}$ & $1.6\times 10^{-2}$ & $1.6\times 10^{-2}$\\ 
 10 & $2.0\times 10^{-2}$ & $2.4\times 10^{-2}$&$3.2\times 10^{-2}$ & $3.2\times 10^{-2}$\\ 
\hline \hline 
\multicolumn{5}{|c|}{$\sqrt{|{\rm{Im}} (\delta^d_{12})^2_{LR}|}$} \\
 \hline 
 2 & $2.7\times 10^{-3}$ & $1.8\times 10^{-3}$ & $2.3\times 10^{-3}$ & $2.1\times 10^{-3}$ \\ 
 5 & $1.3\times 10^{-2}$ & $8.4\times 10^{-3}$ & $1.1\times 10^{-2}$ & $9.4\times 10^{-3}$\\ 
 10 & $4.5\times 10^{-2}$ & $2.9\times 10^{-2}$&$3.7\times 10^{-2}$ & $3.2\times 10^{-2}$\\ 
\hline \hline 
\multicolumn{5}{|c|}{$\sqrt{|{\rm{Im}} (\delta^d_{12})_{LR}={\rm{Im}} (\delta^d_{12})_{RL}|}$} \\ 
\hline
 2 & $8.6\times 10^{-4}$ & $4.5\times 10^{-4}$ & $4.6\times 10^{-4}$ & $3.7\times 10^{-4}$ \\ 
 5 & $2.0\times 10^{-3}$ & $1.0\times 10^{-3}$ & $1.0\times 10^{-3}$ & $8.3\times 10^{-4}$\\ 
 10 & $3.9\times 10^{-3}$ & $2.0\times 10^{-3}$&$2.0\times 10^{-3}$ & $1.6\times 10^{-3}$\\ 
\hline \hline 
\multicolumn{5}{|c|}{$\sqrt{|{\rm{Im}} (\delta^d_{12})^2_{LL}={\rm{Im}} (\delta^d_{12})^2_{RR}|}$} \\ 
\hline
 2 & $1.4\times 10^{-3}$ & $6.7\times 10^{-4}$ & $6.6\times 10^{-4}$ & $5.4\times 10^{-4}$ \\ 
 5 & $5.4\times 10^{-3}$ & $2.1\times 10^{-3}$ & $2.0\times 10^{-3}$ & $1.6\times 10^{-3}$\\ 
 10 & $1.4\times 10^{-2}$ & $4.7\times 10^{-3}$&$4.5\times 10^{-3}$ & $3.6\times 10^{-3}$\\ 
\hline \hline 
\end{tabular} 
\caption{{\it Limits on ${\rm Im }   (\delta^d_{12})_{AB}$ from $\epsilon_K$ with gaugino masses of 250  
 GeV.}}
\label{tab:e250}
\end{center} 
\end{table}

\begin{table} 
\begin{center} 
\begin{tabular}{|c|c|c|c|c|} 
\hline 
$M_{sq}$ [TeV] & No-QCD, VIA & LO-VIA & LO,     $B_i$ & NLO, $B_i$ \\ 
\hline \hline
\multicolumn{5}{|c|}{$\sqrt{|{\rm{Im}} (\delta^d_{12})^2_{LL}|}$} \\
 \hline 
 2 & $5.0\times 10^{-3}$ & $5.9\times 10^{-3}$ & $8.0\times 10^{-3}$ & $7.9\times 10^{-3}$ \\ 
 5 & $1.1\times 10^{-2}$ & $1.3\times 10^{-2}$ & $1.7\times 10^{-2}$ & $1.7\times 10^{-2}$\\ 
 10 & $2.1\times 10^{-2}$ & $2.5\times 10^{-2}$&$3.4\times 10^{-2}$ & $3.3\times 10^{-2}$\\ 
\hline \hline 
\multicolumn{5}{|c|}{$\sqrt{|{\rm{Im}} (\delta^d_{12})^2_{LR}|}$} \\ 
\hline 
2 & $2.0\times 10^{-3}$ & $1.4\times 10^{-3}$ & $1.8\times 10^{-3}$ & $1.6\times 10^{-3}$ \\ 
 5 & $8.3\times 10^{-3}$ & $5.5\times 10^{-3}$ & $7.0\times 10^{-3}$ & $6.2\times 10^{-3}$\\ 
 10 & $2.7\times 10^{-2}$ & $1.8\times 10^{-2}$&$2.2\times 10^{-2}$ & $2.0\times 10^{-2}$\\ 
\hline \hline 
\multicolumn{5}{|c|}{$\sqrt{|{\rm{Im}} (\delta^d_{12})_{LR}={\rm{Im}} (\delta^d_{12})_{RL}|}$} \\
 \hline 
 2 & $1.3\times 10^{-3}$ & $6.0\times 10^{-4}$ & $5.9\times 10^{-4}$ & $4.7\times 10^{-4}$ \\ 
 5 & $2.2\times 10^{-3}$ & $1.1\times 10^{-3}$ & $1.1\times 10^{-3}$ & $9.1\times 10^{-4}$\\ 
 10 & $4.2\times 10^{-3}$ & $2.1\times 10^{-3}$&$2.1\times 10^{-3}$ & $1.7\times 10^{-3}$\\ 
\hline \hline 
\multicolumn{5}{|c|}{$\sqrt{|{\rm{Im}} (\delta^d_{12})^2_{LL}={\rm{Im}} (\delta^d_{12})^2_{RR}|}$} \\ 
\hline 
2 & $1.1\times 10^{-3}$ & $5.8\times 10^{-4}$ & $5.8\times 10^{-4}$ & $4.7\times 10^{-4}$ \\ 
 5 & $4.2\times 10^{-3}$ & $1.9\times 10^{-3}$ & $1.8\times 10^{-3}$ & $1.5\times 10^{-3}$\\ 
 10 & $1.1\times 10^{-2}$ & $4.4\times 10^{-3}$&$4.2\times 10^{-3}$ & $3.4\times 10^{-3}$\\ 
\hline \hline 
\end{tabular} 
\caption{{\it Limits on ${\rm Im }   (\delta^d_{12})_{AB}$ from $\epsilon_K$ with gaugino masses of 500  
 GeV.}}
\label{tab:e500}
\end{center} 
\end{table}

\begin{table} 
\begin{center} 
\begin{tabular}{|c|c|c|c|c|} 
\hline 
$M_{sq}$ [TeV] & No-QCD, VIA & LO-VIA & LO,     $B_i$ & NLO, $B_i$ \\ 
\hline \hline
\multicolumn{5}{|c|}{$\sqrt{|{\rm{Im}} (\delta^d_{12})^2_{LL}|}$} \\ 
\hline 
2 & $7.7\times 10^{-3}$ & $9.0\times 10^{-3}$ & $1.2\times 10^{-2}$ & $1.2\times 10^{-2}$ \\ 
 5 & $1.3\times 10^{-2}$ & $1.5\times 10^{-2}$ & $2.0\times 10^{-2}$ & $2.0\times 10^{-2}$\\ 
 10 & $2.3\times 10^{-2}$ & $2.7\times 10^{-2}$&$3.7\times 10^{-2}$ & $3.6\times 10^{-2}$\\ 
\hline \hline 
\multicolumn{5}{|c|}{$\sqrt{|{\rm{Im}} (\delta^d_{12})^2_{LR}|}$} \\ 
\hline
 2 & $1.8\times 10^{-3}$ & $1.3\times 10^{-3}$ & $1.6\times 10^{-3}$ & $1.4\times 10^{-3}$ \\ 
 5 & $6.0\times 10^{-3}$ & $4.0\times 10^{-3}$ & $5.0\times 10^{-3}$ & $4.5\times 10^{-3}$\\ 
 10 & $1.8\times 10^{-2}$ & $1.1\times 10^{-2}$&$1.5\times 10^{-2}$ & $1.3\times 10^{-2}$\\ 
\hline \hline 
\multicolumn{5}{|c|}{$\sqrt{|{\rm{Im}} (\delta^d_{12})_{LR}={\rm{Im}} (\delta^d_{12})_{RL}|}$} \\ 
\hline
 2 & $5.5\times 10^{-3}$ & $1.3\times 10^{-3}$ & $1.0\times 10^{-3}$ & $7.8\times 10^{-4}$ \\ 
 5 & $2.9\times 10^{-3}$ & $1.4\times 10^{-3}$ & $1.4\times 10^{-3}$ & $1.1\times 10^{-3}$\\ 
 10 & $4.7\times 10^{-3}$ & $2.3\times 10^{-3}$&$2.3\times 10^{-3}$ & $1.9\times 10^{-3}$\\ 
\hline \hline 
\multicolumn{5}{|c|}{$\sqrt{|{\rm{Im}} (\delta^d_{12})^2_{LL}={\rm{Im}} (\delta^d_{12})^2_{RR}|}$} \\ 
\hline 
2 & $1.0\times 10^{-3}$ & $5.5\times 10^{-4}$ & $5.6\times 10^{-4}$ & $4.6\times 10^{-4}$ \\ 
 5 & $3.3\times 10^{-3}$ & $1.6\times 10^{-3}$ & $1.6\times 10^{-3}$ & $1.3\times 10^{-3}$\\ 
 10 & $8.9\times 10^{-3}$ & $3.8\times 10^{-3}$&$3.7\times 10^{-3}$ & $3.0\times 10^{-3}$\\ 
\hline \hline 
\end{tabular} 
\caption{{\it Limits on ${\rm Im }   (\delta^d_{12})_{AB}$ from $\epsilon_K$ with gaugino masses of 1000   GeV.}}
\label{tab:e1000}
\end{center} 
\end{table} 
}


\end{document}